\documentclass[nobm]{epl}

\usepackage{amsmath}

\title{Analytical approximation for reaction-diffusion 
processes in rough pores}
\shorttitle{Analytical approximation...}

\author{J. S. Andrade Jr.\inst{1,3} \thanks{E-mail:
soares@fisica.ufc.br}
\and M. Filoche\inst{2} \thanks{E-mail:
marcel.filoche@polytechnique.fr}
\and B. Sapoval\inst{1,2} \thanks{E-mail:
bernard.sapoval@polytechnique.fr}
}

\institute{
\inst{1} Centre de Math\'ematiques et de leurs
Applications - Ecole Normale Sup\'erieure de Cachan, 94235 Cachan,
France\\
\inst{2} Laboratoire de Physique de la Mati\`ere Condens\'ee
- Ecole Polytechnique, 91128 Palaiseau, France\\
\inst{3} Departamento
de F\'\i sica, Universidade Federal do Cear\'a, - 60451-970 Fortaleza,
Cear\'a, Brazil }


\begin{document}

\maketitle

\begin{abstract}
The concept of an active zone in Laplacian transport is used to obtain
an analytical approximation for the reactive effectiveness of a pore
with an arbitrary rough geometry. We show that this approximation is
in very good agreement with direct numerical simulations performed
over a wide range of diffusion-reaction conditions (i.e., with or
without screening effects). In particular, we find that in most
practical situations, the effect of roughness is to increase the
intrinsic reaction rate by a geometrical factor, namely, the ratio
between the real and the apparent surface area. We show that this
simple geometrical information is sufficient to characterize the
reactive effectiveness of a pore, in spite of the complex
morphological features it might possess.
\end{abstract}

The physics of diffusion-reaction phenomena in disordered media has
attracted a lot of attention in recent years, due to its close
connections with major technological subjects like catalysis and waste
disposal. The development of modeling techniques for the description
of heterogeneous catalytic processes represents a real challenge,
mainly due to the limitations of classical pseudo-homogeneous
representations \cite{Froment90,Thomas97}.  These macroscopic models
for diffusion-reaction processes can only implicitly account for the
geometrical features of real pore spaces \cite{Sahimi95}. Due to {\it
screening effects}, even if the active sites are uniformly distributed
in space along the entire surface, the regions corresponding to fins
or extended protrusions will display a higher activity as opposed to
the deep parts of fjords which are more difficult to access. We recall
that the standard modeling approach is to consider the catalyst
particle as a homogeneous system where reagents and products can
diffuse and react according to a given effective diffusion coefficient
and an intrinsic reaction mechanism. Although in recent years some
effort has been dedicated to the investigation of diffusion and
reaction in heterogeneous geometries (e.g., fractals)
\cite{Avnir89,Gutfraind92,Andrade97,Gavrilov97}, the role played by
the pore shape or the local surface morphology on the overall
diffusivity and reactivity of the catalyst remains to be understood.

The notion of {\it active zone} \cite{Sapoval94} has been recently
applied to elucidate several relevant features of the Laplacian
transport to and across irregular interfaces (e.g., electrodes and
membranes). In the present letter, we show that this concept can be
used to derive an analytical approximation for the diffusion-reaction
activity of pores of arbitrary rough geometry. In a first step, we use
a coarse-graining argument to obtain a general expression for the {\it
effective} reaction rate of an irregular wall element. In the second
step, we apply this result to calculate the reactive effectiveness of
a deep rough pore analytically. Finally, this two-step approach is
validated by direct numerical simulations of the full problem.

\begin{figure}
\onefigure[scale=0.3]{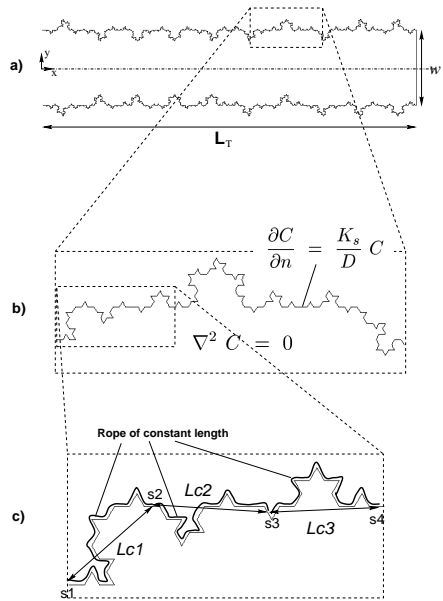}
\caption{(a) Schematic representation of a two-dimensional rough pore.
The reactive walls are composed of 10 successive different random Koch
curves (RKC) of the third generation. (b) Schematic representation of
the wall element. The reactive interface is irregular and the reactant
concentration obeys the Laplace equation. A constant concentration
$C_{S}$ is maintained at the source line and the boundary condition at
the interface is $\partial C / \partial n = K_{s} C/D$. (c) Schematic
representation of the coarse-graining or finite renormalization
procedure used in the ``Land Surveyor Approximation''
(LSA). Successive ropes of length $\Lambda$ determine successive
chords of lengths $L_{c1}$, $L_{c2}$,...}
\label{fig:1} 
\end{figure}

As shown in Fig.~\ref{fig:1}, we consider the linear and steady state
diffusion transport of reactant $A$ in a single slit-shaped pore of
nominal length $L_T$ ($x$-direction). The two reactive walls of this
pore are rough surfaces of thickness $b$ ($z$-direction) separated by
a distance $w$ ($y$-direction). By definition, they are generated
through the translation along the $z$-direction of two lines made by
the connection of several elementary irregular interfaces. In the
example shown in Fig.~\ref{fig:1}(a), the walls of the pore have a
fractal geometry of a 3-generation random Koch curve (RKC)
\cite{Mandelbrot82,Filoche00}. From a source at the pore entrance,
molecules of species $A$ diffuse into the pore through a neutral
solvent to react at the walls according to a first-order kinetics, $A
\rightarrow A^{\star}$.
  
The basic idea is to focus on a typical finite region of the pore
enclosing a wall that corresponds to a single elementary unit of the
interface. This subset of the system can be approximately treated as a
diffusion-reaction cell where reagent species $A$ diffuse from a
source line of length $L$ towards an irregular reactive interface of
perimeter $L_p$ and thickness $b$ (see Fig.~\ref{fig:1}(b)). In the
bulk of this cell, the diffusion of species $A$ obeys Fick's law,
$\vec{J}(\vec{r}) \equiv -D \vec{\nabla} C$, where $\vec{J}$
represents the mass flux vector field, $C(\vec{r})$ is the local
concentration at position $\vec{r}$ and $D$ is the molecular diffusion
coefficient. Under steady state conditions, the concentration field of
the reactant species satisfies the stationary diffusion equation
${\nabla}^{2} C=0$. For the boundary conditions, we assume a constant
concentration of $A$ equal to $C_S$ at the source line. At the
reactive interface, the depletion rate of $A$ follows $J_n \equiv -
K_{s}C$, where $K_{s}$ is the {\it intrinsic} reaction rate
constant. Due to mass conservation, the local reaction flux must be
equal to the diffusive flux reaching the active interface from the
bulk. As a consequence, the boundary condition at this point can be
written as

\begin{equation} 
\frac{\partial C}{\partial n} = \frac{K_{s}}{D}~C~~~. \label{eq:fourier} 
\end{equation} 

This introduces a finite length scale $\Lambda \equiv D/K_{s}$ into
the problem. Thus, in order to understand the diffusion-reaction
behavior of irregular interfaces, one has to solve the Laplace
equation with the boundary condition (\ref{eq:fourier}) on the
walls. This is an unsolved mathematical problem.

In contrast, the problem of the Dirichlet boundary condition ($C=0$)
on irregular surfaces has been thoroughly studied, specially for the
case of two-dimensional systems. An important theorem proposed by
Makarov \cite{Makarov85} has been used to describe the properties of
the current distribution on irregular electrodes (e.g, fractal
electrodes) \cite{Sapoval94}. This theorem states that {\it the
information dimension of the harmonic measure on a singly connected
electrode in $d=2$ is exactly equal to 1}. In terms of activity, this
means that, whatever the shape of the interface, the size $L_{act}$ of
the region where most of the reaction takes place is of the order of
the overall size (or diameter) $L$ of the cell under a dilation
transformation. When applied to a transport system like the
diffusion-reaction cell shown in Fig.~\ref{fig:1}(b), Makarov's
theorem has a simple but general consequence: the screening effect due
to geometrical irregularities of the reactive interface can be
characterized in terms of the ratio \cite{Sapoval94}

\begin{equation}
S \equiv L_{p}/L~~~.
\end{equation}

Because the active length of the interface is of the order of the
system size, $L_{act} \approx L$, then $L_{act} \approx L_{p}/S$ and
the factor $S$ can be considered as the ``screening factor'' of the
Dirichlet-Laplacian field. This result cannot be applied as such to a
real diffusion-reaction cell because the boundary condition on the
reacting interface is not $C=0$, but rather $\partial C /\partial n =
C/\Lambda$. To account for this more realistic boundary condition, we
use the Land Surveyor Approximation (LSA)
\cite{Sapoval94,Filoche97,Sapoval99}. In this method, one switches
from the real geometry obeying the real boundary condition to a
coarse-grained geometry obeying the Dirichlet boundary condition,
where the coarse-graining depends on $\Lambda$. The size of the active
zone of the reactive interface can then be obtained from the screening
factor of that new geometry.

Consider a region $i$ along the reactive interface with perimeter
$L_{pi}=s_{i+1}-s_{i}$, where $s$ is the curvilinear coordinate along
the interface. It has a reactive conductance $G_{i}=[bK_{s}] L_{pi}$.
The access conductance to reach this surface is $G_{\rm acc} \approx
bD$. Depending on the size of the region $i$, there exists two
situations: $G_{i} < G_{\rm acc}$ or $G_{\rm acc} < G_{i}$. If
$L_{pi}$ is small, $G_{i} < G_{\rm acc}$ and the flux is limited by
$G_{i}$. If $L_{pi}$ is large enough, the flux is limited by the
access conductance. In the latter situation we are, in a first
approximation, back to the case of a Laplacian field with Dirichlet
boundary condition (i.e., $C=0$). The new geometry is defined through
the identity $G_{i}=G_{\rm acc}$, which means that the curvilinear
distance $L_{pi}$ along the reactive surface is equal to
$\Lambda$. From its definition, a coarse-grained region can be
considered as acting uniformly. If we performed the coarse-graining on
a larger scale, the flux would no longer be uniform within a macrosite
and, therefore, we would not be able to identify the size of the
active zone. The size of a grain, $L_c$, is the distance in real space
(or the chord length) between $s_1$ and $s_2$. We then shift from the
real geometry to the coarse-grained geometry which is made of
successive chords $L_{c1} = L_c(s_0,s_1)$, $L_{c2} = L_c(s_1,s_2)$,
\ldots~(see Fig.~\ref{fig:1}(c)).
 
If we now observe that the perimeter of the coarse-grained interface
is $L_{p,cg}=L_{c1}+L_{c2}+...= N \langle L_{c}\rangle$, where
$\langle L_{c}\rangle$ is the average chord length, it is possible to
express the screening factor of the coarse-grained geometry as
$S_{cg}=N \langle L_{c}\rangle /L$. In this way, the effective
conductance of the interface can be written as $G_{\rm
eff}=bL_{p}K_{s}/S_{cg}$. By definition, since each chord corresponds
to a perimeter $\Lambda$, we can write $L_{p}=N \Lambda$ and then
express the conductance of the reactive interface as \cite{Sapoval99}

\begin{equation} 
G_{\rm eff}=bD~\frac{L}{\langle L_c \rangle}~~~. \label{eq:cond} 
\end{equation} 

This result expresses how the conductance of a typical wall element
should depend on the diffusivity of the reagent species in the bulk
phase and the average chord length corresponding to a perimeter of
length $\Lambda$. The geometry enters through the local relation
between a perimeter of length $\Lambda$ and its associated chord
length. At this point, we can then define from (\ref{eq:cond}) an {\it
effective} reaction rate constant for the interface as

\begin{equation} 
K_{\rm eff}=D/\langle L_{c}\rangle~~~. \label{eq:react} 
\end{equation} 

This result is general. It applies to an arbitrary wall geometry
provided that the wall itself does not contain deep pores
\cite{Sapoval99}. In the particular case where the wall irregularity
can be described by a fractal geometry, the chord length $\langle L_c
\rangle$ of a self-similar interface with dimension $D_f$, size $L$
and lower cut-off length scale $\ell$ can be expressed approximately
as

\begin{equation} 
\langle L_{c}\rangle= 
\begin{cases} 
\Lambda & \text{if $\Lambda \ll \ell$},\\ 
\ell {(\Lambda /\ell)}^{\frac{1}{D_f}} & 
\text{if $\ell < \Lambda < L_p$},\\ 
\Lambda/S & \text{if $\Lambda \gg L_p$}~~~. 
\end{cases}  
\label{eq:lameff} 
\end{equation} 

Note that in the case where $\Lambda \gg L_p$, Eq.~(\ref{eq:react}) becomes

\begin{equation} 
K_{\rm eff}=SK_{s}~~~. \label{eq:reactS} 
\end{equation} 

This exact relation provides a very simple expression for the effect 
of the roughness on the effective reaction rate. 

In the second step, the LSA result (\ref{eq:react}) is used to compute
the overall activity of the rough pore by considering it as a smooth
pore with an effective reactivity. Thus, writing this reactivity of
the coarse-grained wall in terms of $K_{\rm eff}$, it is possible to
describe by mass conservation the phenomenon of diffusion and surface
reaction over the entire pore using the following differential
equation:

\begin{equation} 
{\partial^2C\over\partial x^2} + {\partial^2C\over \partial y^2}=0
\label{eq:dif2d}
\end{equation} 

with boundary conditions

\begin{align} 
C(0,y)=C_{S}~~,~~~{\partial C\over\partial x}(L_{T},y)=0~~,~~~
{\partial C\over\partial y}(x,0)=0 \notag \\ \text{and}~~~~~~{\partial
C\over\partial y}(x,w/2)+\frac{K_{\rm eff}}{D}C(x,w/2)=0~~~.
\label{eq:bc2d} 
\end{align} 

The boundary condition ${\partial C\over\partial y} (x,0)=0$
corresponds to the assumption that the pore is symmetric with respect
to $y=0$. As in heterogeneous catalysis studies, the effect of
reaction-diffusion process is measured by an effectiveness factor
\cite{Thomas97}:

\begin{equation} 
\eta \equiv \frac{\Phi}{\Phi_{\rm max}}~~~, \label{eq:effec}
\end{equation} 

where $\Phi$ is the mass flux penetrating the system 

\begin{equation} 
\Phi=2b\int_{0}^{w/2}\biggl[-D{\partial C\over\partial
x}(0,y)\biggr]dy~~~,
\end{equation} 

and $\Phi_{\rm max}=2DbL_{T}SC_{S}/\Lambda$ is the mass flux in the
absence of diffusion limitations. From the analytic solution of
Eqs.~(\ref{eq:dif2d}) and (\ref{eq:bc2d}) and using the definition
(\ref{eq:effec}), we obtain the following expression for the reactive
effectiveness of a two-dimensional rough pore \cite{Carslaw59}:

\begin{equation} 
\eta=\frac{2\Lambda}{\langle L_{c}\rangle L_{T}S} 
\sum_{n=1}^{\infty}\frac{\tan(\alpha_{n}w/2) 
\tanh(\alpha_{n}L_{T})}{[(\alpha_{n}^2+{\langle L_{c}\rangle}^{-2})w/2 
+{\langle L_{c}\rangle}^{-1}]} \label{eq:effec2d} 
\end{equation} 

where $\alpha_{n}$ is the $n$th root of the eigenvalue equation,
$\alpha_{n} \tan(\alpha_{n} w/2)={\langle L_{c}\rangle}^{-1}$ for
$n=1,2,3,\dots$

\begin{figure}
\onefigure[scale=0.3]{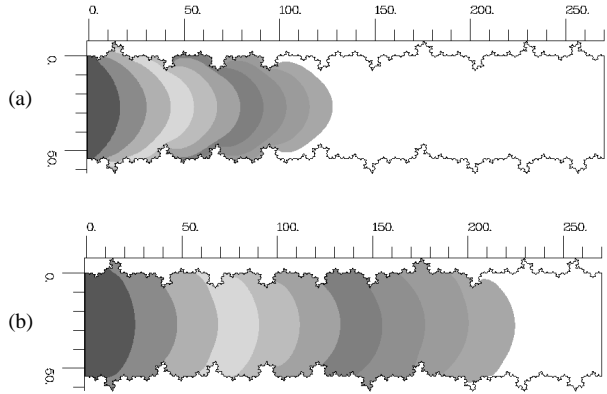}
\caption{Contour plot of the steady-state concentration of reactant
penetrating a rough pore. The relevant sizes of the pore are $w=2L$
and $L_{T}=10L$. In (a) $\Lambda/\ell=10$ while in (b)
$\Lambda/\ell=100$. The contour plots have been computed from direct
numerical simulation using the finite elements technique. The change
in the grey scale corresponds to a factor of 2 in concentration.}
\label{fig:2} 
\end{figure}

In order to check the validity of Eqs.~(\ref{eq:lameff}) and
(\ref{eq:effec2d}), we numerically calculate the solution of the
two-dimensional Laplacian transport in a rough pore geometry whose
active walls are composed by several elementary RKC interfaces
($D_{f}=\log 4/\log 3$). This task is performed here through
discretization with unstructured meshes by means of the finite
elements technique \cite{Modulef}. Due to computational limitations,
we restrict our simulations to reactive pore walls made by RKC curves
of $2$ and $3$ generations. In Figs.~\ref{fig:2}(a) and (b), we show
the concentration fields in a typical reactive rough pore constructed
with $3$-generations RKC walls and calculated at $\Lambda/\ell=10$ and
$100$, respectively.

\begin{figure}
\onefigure[scale=0.4]{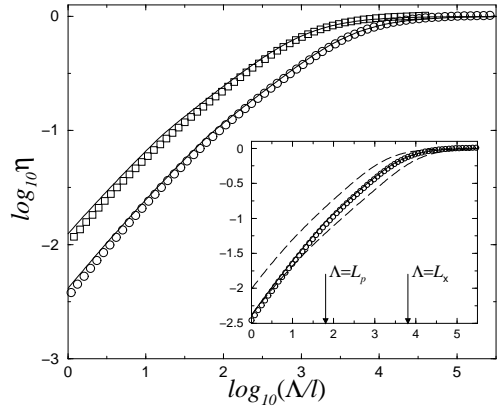}
\caption{Dependence of the effectiveness factor $\eta$ on
$\Lambda/\ell$. The symbols correspond to $\eta$ values calculated
from direct numerical simulation of reactive rough pores whose walls
are made by RKC's of second (squares) and third (circles)
generation. The thick solid lines correspond to the analytical
approximations from Eqs.~(\ref{eq:lameff}) and (\ref{eq:effec2d}). The
arrows point to the crossovers $\Lambda=L_{p}$ and $\Lambda=\
L_{\times}$. For comparison, the inset shows two curves corresponding
to the effectiveness behavior of smooth pores (dashed lines) with
lengths $L_{T}$ (top) and $L_{T}S$ (bottom).}
\label{fig:3} 
\end{figure}

The results displayed in Fig.~\ref{fig:3} show that the LSA
predictions for the effectiveness are in excellent agreement with
numerical simulations of the complete diffusion-reaction equations in
both $2$- and $3$-generations RKC pores and over the whole range of
relevant $\eta$ variability, namely, for $\Lambda>\ell$
\cite{regime}. It is illustrative to compare the results obtained for
a rough pore with the classical behavior of reactive, but flat pores.
In the inset of Fig.~\ref{fig:3}, we show the logarithmic curves of
$\eta$ versus $\Lambda/\ell$ calculated for a rough pore
($3$-generations RKC) and for the corresponding smooth pores with
lengths $L_{T}$ and $L_{T}S$. The large discrepancy which can be
observed among the behaviors of rough and flat pores clearly indicates
the important role played by the geometry of the reactive
interfaces. Finally, one should note that, in the limit of negligible
resistance to pore diffusion, the effectiveness approaches
asymptotically its maximum, $\eta=1$. If the pore is very deep ($L_{T}
\gg w$), the one-dimensional approximation of Eq.~(\ref{eq:dif2d})
allows one to estimate the crossover for this saturation regime as
$\Lambda=L_{\times} \equiv 2{L_{T}}^2S/w$. This prediction is also in
good agreement with the numerical simulations (see the inset of
Fig.~\ref{fig:3}).

In summary, the work presented here provides clear evidence that the
effect of pore space morphology on the global efficiency of a
diffusion-reaction system is not only relevant, but should govern the
reactivity of irregular interfaces under diffusion limitations. In
particular, it is found that, contrary to the classical description
\cite{Froment90,Thomas97}, the effect of an arbitrary rough geometry
is to modify the reaction rate and {\it not} the effective diffusion
coefficient. Note that the analytical approximation derived here and
validated through direct numerical simulations rely solely on the
knowledge of simple geometrical characteristics of the interface to
provide accurate predictions for the effectiveness of rough pores,
over a wide range of diffusion-reaction conditions. For all practical
situations where the reactant penetration in the pore is significant,
one can use the simple model of a smooth pore with an effective
reactivity $K_{\rm eff}$ to describe the efficiency of a rough
pore. Remarkably, $K_{\rm eff}$ is the product of the intrinsic
reactivity by the screening factor $S$, which has an elementary
geometrical meaning.

At last, since the Land Surveyor Method has also been successfully
applied to the response of non-linear irregular electrodes
\cite{Filoche00}, it could possibly be used to study higher order
reactions in rough pores.

\acknowledgments

This work has been partially supported by CNPq.


\begin{thebibliography}{0}
\bibitem{regime}
{The regime $\Lambda<\ell$ corresponds to a highly
reactive pore wall ($K_{s}\rightarrow\infty$) where the reactant
penetration is negligible. This regime is only interesting from a
theoretical point of view because real diffusion-reaction systems
should not operate under significant mass transport limitations.}
\end{thebibliography}
\end{document}